\documentclass[a4paper,11pt]{article}
\pdfoutput=1 % if your are submitting a pdflatex (i.e. if you have
\usepackage{jinstpub} % for details on the use of the package, please
\usepackage{color}% for personal comment
\usepackage{float}% for floating figures

\title{\boldmath Development of LGAD sensors with a thin entrance window for soft X-ray detection}

\author[a,1]{Jiaguo Zhang,\note{Corresponding author.}}
\author[a]{Rebecca Barten,}
\author[a]{Filippo Baruffaldi,}
\author[a]{Anna Bergamaschi,}
\author[b]{Giacomo Borghi,}
\author[b]{Maurizio Boscardin,}
\author[a]{Martin Br\"uckner,}
\author[a]{Maria Carulla,}
\author[b]{Matteo Centis Vignali,}
\author[a]{Roberto Dinapoli,}
\author[a]{Simon Ebner,}
\author[b]{Francesco Ficorella,}
\author[a]{Erik Fro\"jdh,}
\author[a]{Dominic Greiffenberg,}
\author[b]{Omar Hammad Ali,}
\author[a]{Julian Heymes,}
\author[a]{Shqipe Hasanaj,}
\author[a]{Viktoria Hinger,}
\author[a]{Thomas King,}
\author[a]{Pawel Kozlowski,}
\author[a]{Carlos Lopez-Cuenca,}
\author[a]{Davide Mezza,}
\author[a]{Konstantinos Moustakas,}
\author[a]{Aldo Mozzanica,}
\author[b]{Giovanni Paternoster,}
\author[b]{Sabina Ronchin,}
\author[a]{Christian Ruder,}
\author[a]{Bernd Schmitt,}
\author[a]{Dhanya Thattil}

\affiliation[a]{Paul Scherrer Institut,\\5232 Villigen, Switzerland}
\affiliation[b]{Fondazione Bruno Kessler,\\Via Sommarive 18, 38123 Trento, Italy}

\emailAdd{jiaguo.zhang@psi.ch}

\abstract{
We show the developments carried out to improve the silicon sensor technology for the detection of soft X-rays with hybrid X-ray detectors. An optimization of the entrance window technology is required to improve the quantum efficiency. The LGAD technology can be used to amplify the signal generated by the X-rays and to increase the signal-to-noise ratio, making single photon resolution in the soft X-ray energy range possible. In this paper, we report first results obtained from an LGAD sensor production with an optimized thin entrance window. Single photon detection of soft X-rays down to 452~eV has been demonstrated from measurements, with a signal-to-noise ratio better than 20.
}


\begin{document}
\maketitle
\flushbottom

%\linenumbers

\section{Introduction}
\label{sec:intro}

Soft X-ray applications at synchrotrons and FELs are limited by the performance of the currently available detectors using silicon sensors. The main issues are their low quantum efficiency (QE) due to the shallow absorption of the photons in the entrance window of the sensor, and their difficulties in achieving single photon resolution, since the small amount of charge generated by the low energy X-rays is often comparable to the electronic noise. At present, the hybrid X-ray detector technology is widely used for hard X-ray experiments thanks to the high frame rate, large dynamic range, large sensitive area, stability and robustness~\cite{Anna2020}. Further development of the hybrid X-ray detectors with equivalent performance in the soft X-ray energy range would be beneficial for several diffraction, spectro-microscopy and imaging experiments which have to be performed at low energies due to the low interaction of the sample with the radiation or the presence of characteristic edges interesting for research~\cite{Hitchcock2015}. Examples include Resonant Inelastic X-ray Scattering (RIXS) experiments at FELs and synchrotrons, which requires single photon resolution down to 250~eV as well as high spatial resolution, and magnetic contrast imaging at the L-edges of 3d transition metals using ptychography and absorption spectroscopy~\cite{Rafael2018}.

In collaboration with the sensor manufacturer Fondazione Bruno Kessler (FBK), the Paul Scherrer Institut (PSI) is developing and optimizing the Low Gain Avalanche Diode (LGAD) sensor technology as well as the Thin Entrance Window (TEW) technology targeting soft X-rays. The TEW technology enables the improvement of the QE for soft X-rays and the LGAD technology increases the signal amplitude and signal-to-noise ratio (SNR), making single photon resolution possible in this energy range. However, LGAD is a technology first introduced in high energy physics because of its timing performance and still needs to be optimized for soft X-ray applications.

The paper is structured as follows: In section~\ref{sec:limitation}, the limitation of hybrid X-ray detectors for soft X-ray detection will be discussed in more details. Section~\ref{sec:strategies} will introduce the choices and strategies of this development. The first results with soft X-ray photons down to 452 eV using the developed LGAD sensor combined with an optimized TEW will be shown in section~\ref{sec:results}. Finally, we will summarize the on-going work and give an outlook for future developments in section~\ref{sec:summary}. 

\section{Limitations of hybrid detectors for soft X-rays}
\label{sec:limitation}

The current limitations of hybrid detectors for soft X-ray detection are mainly due to their low QE and the electronic noise.

\subsection{Reduced QE}
%\delete{QE loss}\add{subsection require too much space}

\begin{figure}
\centering
(a)\includegraphics[height=47mm]{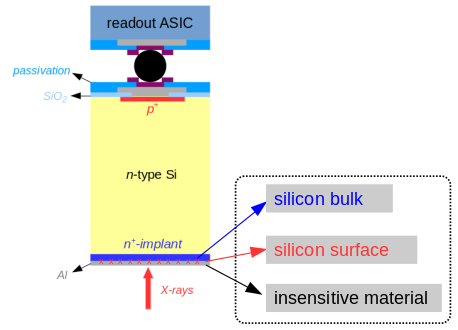}
(b)\includegraphics[height=47mm]{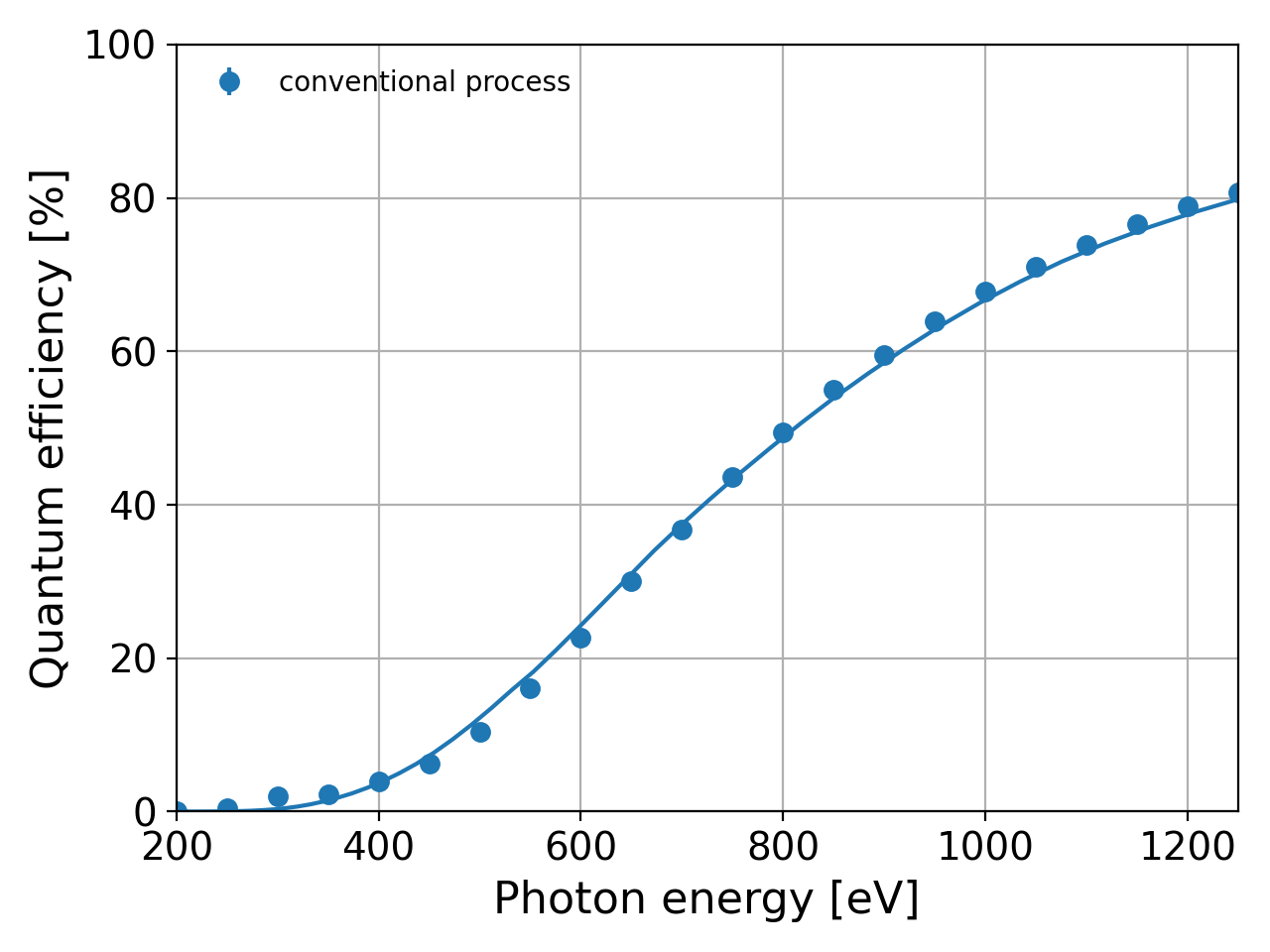}
\caption{(a) Cross section of a conventional silicon sensor showing the locations where the QE loss occurs. (b) The measured (dots) and simulated (line) QE for a conventional silicon sensor consisting of a deep $n^{+}$-implant of $\sim$ 1-2 $\mu$m and a layer of aluminum of $\sim$ 500 nm on the backside of the sensor.}
\label{Cross_section}
\end{figure}

The main reasons for the low QE efficiency for soft X-rays are summarized in figure~\ref{Cross_section}(a)~\cite{Granato2012}:

\begin{itemize}
\item absorption of the photons in the insensitive layer on the sensor backplane;
\item recombination of charge carriers generated by X-rays at the silicon surface;
\item recombination of charge carriers in the highly doped region in silicon. 
\end{itemize}

The absorption of X-ray photons and a partial loss of charge carriers due to recombination result in a low QE and limit the overall detection efficiency for soft X-rays. For example, figure~\ref{Cross_section}(b) shows the measured (dots) and simulated (line) QE as a function of photon energy in the energy range from 200 eV to 1250 eV from a conventional silicon sensor usually used for hard X-ray experiments. At 400~eV the QE is less than 5\%, which is too low for a photon-hungry experimental technique like RIXS. In addition, the incomplete charge collection due to the recombination of charge carriers at the silicon surface as well as in the highly doped region causes bad spectroscopic performance and makes interpolation difficult because part of the charge produced by X-ray photons is lost. 

%\delete{Thus an optimized TEW technology is necessary in this development in order to overcome the QE loss in the soft X-ray energy range.}

\subsection{Electronic noise}
%\delete{Electronic noise}\add{spare some space}

Hybrid X-ray detectors suffer from a relatively high electronic noise, due to the large input capacitance dominated by the inter-electrode capacitance from the silicon sensor as well as the interconnection between the sensor and ASICs (bump-bonding for pixels and wire-bonding for strips). The Equivalent Noise Charge (ENC), defined as the signal at the input equivalent to the electronic noise measured by the detector, is often comparable to the signal produced by soft X-ray photons. This results in a low SNR in the soft X-ray energy range limiting the lowest detectable photon energy. For example, the electronic noise in the Jungfrau and M\"onch detectors using a conventional silicon pixel sensor is $\sim$~35~$e^{-}$~\cite{Viktoria2021, Marco2017}, which allows the detection of photons with an SNR larger than 5 only down to 630~eV (1260~eV with $2\times2$ clustering~\footnote{The electronic noise in a cluster is scaled with the square root of the number of pixels used in the clustering times the noise of a single pixel.})~\cite{Marco2017}.

To overcome the limit due to the electronic noise, we started investigations on LGAD sensors for X-ray detection in 2017. The first results have been reported in~\cite{Marie2019}.  LGADs are able to increase the signal amplitude in the sensor thanks to their internal gain, which takes place by charge multiplication in a high electric field without triggering avalanche breakdown~\cite{Giovanni2017}. This improves the SNR of the detector and allows the detection of the small amount charge of charge created by soft X-rays below 1 keV.

\section{Development choices and strategies}
\label{sec:strategies}

In order to detect soft X-ray photons with high QE and single-photon resolution for hybrid X-ray detectors, the LGAD sensor combined with a TEW is the key in this development. We first optimize the entrance window using planar silicon sensors through different process variations. Once demonstrated, the best TEW process is implemented into the LGAD sensor. Then the LGAD sensor with a TEW is further optimized with an emphasis on the fill factor~\cite{Marie2019} and its gain layer profile. The uniformity of the gain layer due to implantation over a large detection area is another important assessment.

\subsection{TEW optimization}

%\begin{figure}
%\centering
%\includegraphics[width=150mm]{Mechanism.png}
%\caption{Different mechanisms of recombination occuring in the silicon and at the surface.}
%\label{Mechanism}
%\end{figure}

To prevent the complete loss of X-ray photons in the insensitive layer on top of the sensor (typically aluminum for the conventional silicon sensor), it is possible to remove it completely. In case it is required for certain experiments for the purpose of light shielding, for example pump-probe experiments based on a laser stimulus, the thickness of the aluminum has to be minimized without sacrificing the metalization uniformity for a consistent QE over the entire sensor area~\footnote{The thickness of the aluminum in a conventional silicon sensor is typically 1-2 $\mu$m.}.

The partial loss of charge carriers is caused by various kind of recombination. The recombination mainly occurs through the following mechanisms~\cite{SMSze2007}:%, as shown in figure~\ref{Mechanism}:

\begin{itemize}
\item Band-to-band radiative recombination: It is important for direct band semiconductors, $e.g.$ GaAs, but negligible for indirect band semiconductors like silicon; 
\item Shockley-Read-Hall (SRH) recombination (also known as trap-assisted recombination): It occurs due to the existence of traps in the silicon and interface traps at the surface of the silicon;
\item Auger recombination: It becomes dominant for doping concentrations higher than \mbox{$\sim$ 10$^{19}$~cm$^{-3}$}.
\end{itemize}

The SRH and Auger recombinations in the silicon, dominated by the latter one in most of cases, reduce the lifetime of charge carriers and their diffusion length. Through TCAD simulation, the diffusion length for the minority carriers, e.g. holes in the $p^{+}$-on-$n$ sensor, is $\sim$200~nm for a doping concentration of up to $10^{20}$~cm$^{-3}$. Therefore, we are targeting a doping profile with a depth of a couple of hundred nanometers and a peak concentration of $\leq$ $10^{20}$~cm$^{-3}$ through process optimization. With such an implant depth and peak concentration, most of the minority carriers generated by X-ray photons can diffuse out of the low electric field region (highly doped region) and finally be collected by the readout electrodes.

In addition, to reduce the recombination of charge carriers at the silicon surface, where the periodic lattice ends, a thin passivation layer is introduced. The recombination rate at the surface of native silicon can be as high as $\geq$~$10^{7}$~cm/s, which corresponds to a carrier lifetime of $\ll$ 1~ns, whereas the passivated surface, depending on the technology chosen in the process, is able to effectively reduce the surface recombination rate down to 10-100~cm/s.

\begin{figure}
\small
\centering
(a)\includegraphics[width=70mm]{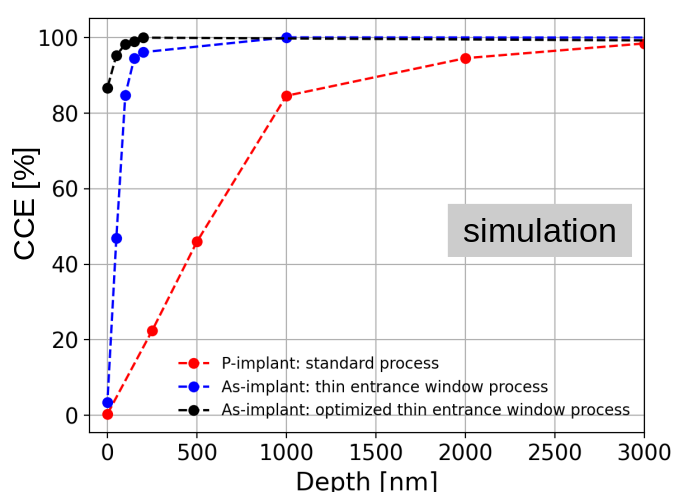}
(b)\includegraphics[width=70mm]{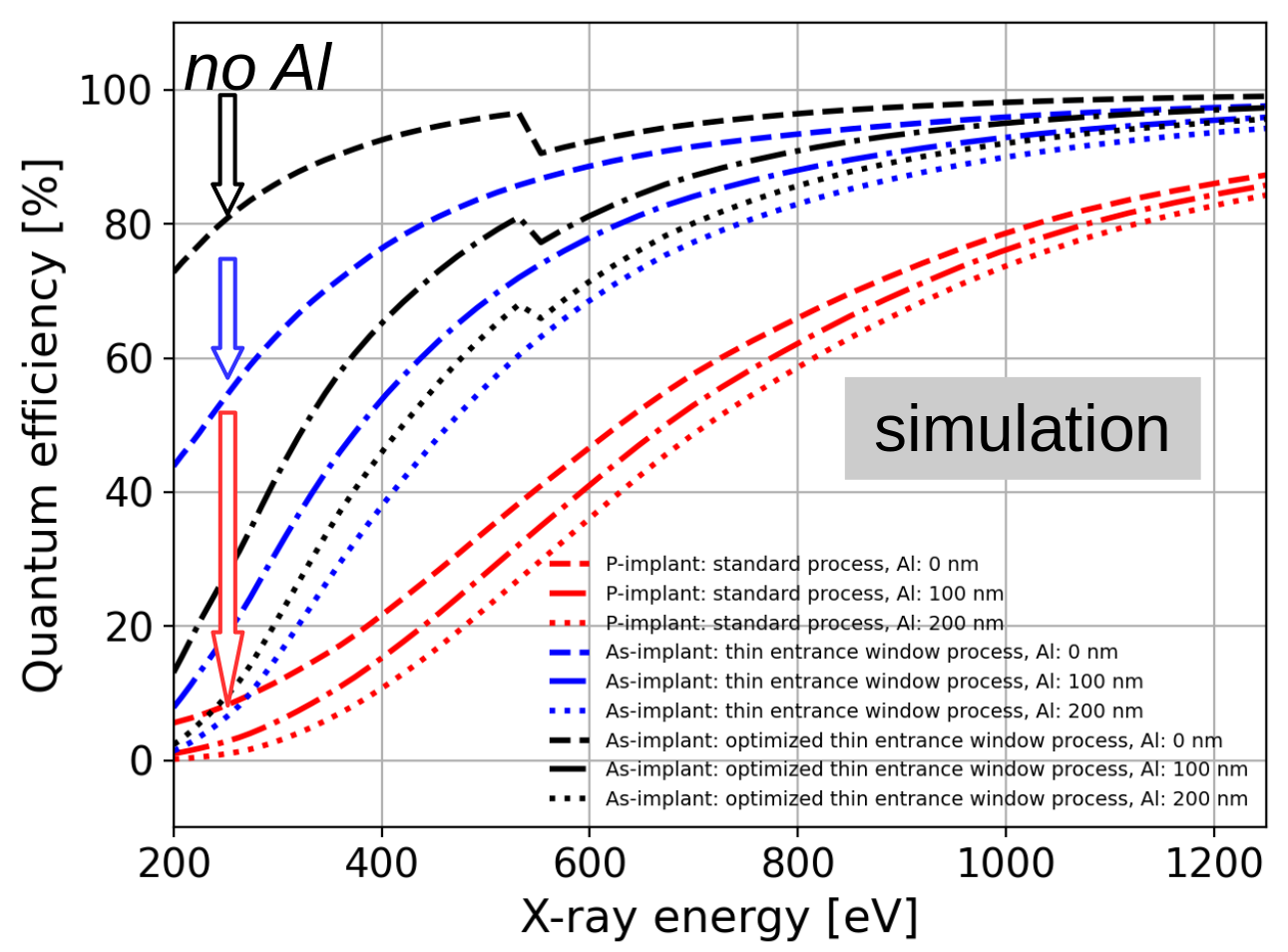}
\caption{(a) Simulated charge-collection efficiency (CCE) vs. depth where X-ray photons are absorbed. (b) Simulated QE vs. X-ray energy for different entrance window options.}
\label{QE_CCE}
\end{figure}

Figure~\ref{QE_CCE}(a) shows the simulated Charge-Collection Efficiency (CCE, defined as percentage of charge collected by the readout electrode) as function of depth where X-ray photons are absorbed for a conventional process using a phosphorus implant (in red), a TEW process using an arsenic implant without passivated surface (in blue), and the optimized TEW process using an arsenic implant with a passivated surface (in black). The optimized TEW process shows much higher CCE ($\geq$~85\%) within the first few nanometer comparing to the others. Using the simulated CCE versus depth, the QE as a function of X-ray energy has been calculated taking the X-ray absorption in different layers into account (aluminum and passivation in case they are present) and shown in figure~\ref{QE_CCE}(b). From the simulation, it is shown that the optimized TEW process combining a shallow implant profile of $\sim$200~nm and a passivated surface is able to achieve a QE of $\geq$~80\% at 250~eV.

\subsection{LGAD optimization}

LGAD technologies can be mainly divided in two different classes: conventional design with the gain layer beneath the segmented readout electrodes, or the inverse-LGAD (iLGAD) design with the gain layer located at the backside of the sensor. 

The first is based on the standard process and its multiplication gain is independent on the energy of soft X-rays since they are absorbed close to the backside surface. However, the segmented gain layer below the readout electrodes of the conventional LGADs results in a limited fill factor, making the interpolation through charge sharing impossible for experiments requiring high spatial resolution. Recent developments, for example the Trench Isolated LGAD (TI-LGAD), AC-coupled LGAD (AC-LGAD) and Deep Junction LGAD (DJ-LGAD) have shown very good progress in this respect. However, they need further developments in order to be used in soft X-ray experiments~\cite{Giovanni2020, Nicolo2020, Ayyoub2021}.

iLGADs with a gain layer covering the entire detection area show 100\% fill factor, making interpolation possible. However, the double-sided process is further complicated by the implementation of a TEW combined with the gain layer on the backplane of the sensor. Moreover, since the the gain layer is located at the entrance window, the multiplication gain depends on the depth where soft X-ray photons are absorbed.

We decided to optimize the iLGAD with a TEW for soft X-ray applications because the iLGAD technology is the most promising choice for imaging experiments in particular those requiring high spatial resolution through interpolation. If the gain layer of the iLGAD is located further away from the surface, the majority of charge carriers are expected to be generated in the $n^{+}$-implant layer and holes will drift through the entire charge multiplication region. In this case, the multiplication gain is independent from the energy of the soft X-rays. However, due to the difference in the impact-ionization coefficients, the multiplication gain will be at least 2-3 times smaller compared to the one of electrons drifting through the same multiplication region. On the other hand, with a gain layer as shallow as possible letting the majority of X-ray photons pass through, the generated electrons will travel through the multiplication region and result in a higher gain in this case. However, soft X-ray photons still have certain probability to be absorbed in the gain layer, resulting in gain variations that depend on the depth where the X-ray photon is absorbed.

\begin{figure}
\small
\centering
(a)\includegraphics[height=50mm]{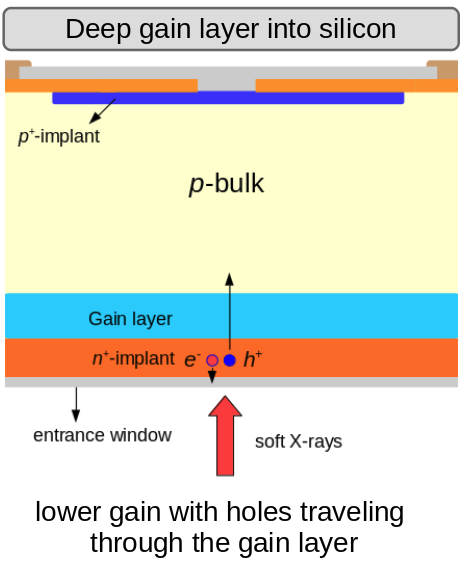}
(b)\includegraphics[height=50mm]{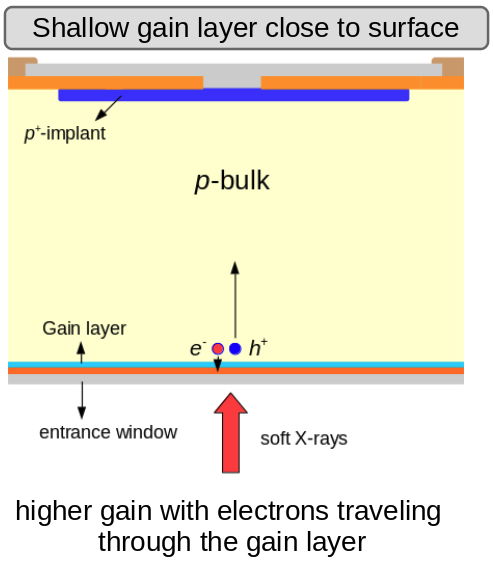}
(c)\includegraphics[height=50mm]{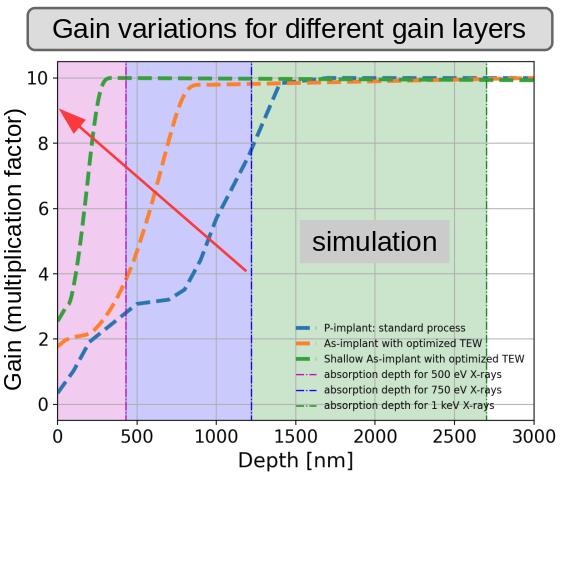}
\caption{(a) A deep gain layer design in the iLGAD with a lower gain due to holes traveling through the gain layer. (b) A shallow gain layer design in the iLGAD with a higher gain due to electrons traveling through the gain layer. (c) Simulated multiplication gain as a function of X-ray absorption depth for different gain layer designs.}
\label{gain_variation}
\end{figure}

Figure~\ref{gain_variation}(a) and (b) shows the the two different choices for the iLGAD technologies and the dependence of the gain on the X-ray absorption depth. Taking advantage of the higher gain from the charge multiplication induced by electrons and thus higher SNR for hybrid X-ray detectors, the iLGADs have been optimized targeting a shallow gain layer towards the technological limit. In figure~\ref{gain_variation}(c), the simulated gain is plotted as a function of depth for the combinations of deep phosphorus and boron implantation (in blue, without TEW), standard arsenic and boron implantation (in orange, with optimized TEW) and the optimized shallow arsenic and boron implantation (in green, with optimized TEW). The attenuation length for 500~eV, 750~eV and 1~keV X-ray photons are indicated as purple, blue and green vertical lines in the figure. According to the simulation, the majority of X-ray photons in this energy range is able to pass through the gain layer of the optimized iLGAD sensor and to result in a higher gain.

\section{Results}
\label{sec:results}

A batch of planar silicon sensors for the optimization of the TEW process has been fabricated at FBK. Measurements of the QE have been performed using UV light of 405~nm and soft X-rays from 200~eV to 1250~eV. The manuscript summarizing the results is in preparation and to be published in separate papers~\cite{Mar2022, Matteo2022}. In this section, we will focus on the first results of the optimized iLGAD sensors with a TEW from the R\&D batch that was produced at FBK. The results to be discussed below are from one process variation; the iLGAD sensors from the other variations will be investigated in details in the future.

A pixelated iLGAD sensor, consisting of 400 $\times$ 400 pixels with a pitch of 25 $\mu$m, has been bump-bonded to the M\"onch-03 charge-integrating readout ASIC for investigations~\footnote{Investigations using another iLGAD sensor with pixels of 75 $\mu$m and Jungrau readout ASIC have been done as well. The results from Jungfrau are discussed in details in a separate paper~\cite{Viktoria2022}.}. Details regarding to the M\"onch-03 readout ASIC can be found in~\cite{Marco2017}. 

Due to the presence of the gain layer in the sensor, the leakage current is multiplied as well compared to conventional planar silicon sensors. The high leakage current results in a reduction of the usable exposure time of charge-integrating detectors in order to avoid saturation of the readout ASIC. This in turn results in a larger shot noise.%The high leakage current shortens the usable exposure time of the detector to avoid saturation of the readout ASIC due to the leakage current and to reduce the shot noise. 

\begin{figure}
\small
\centering
(a)\includegraphics[height=48.5mm]{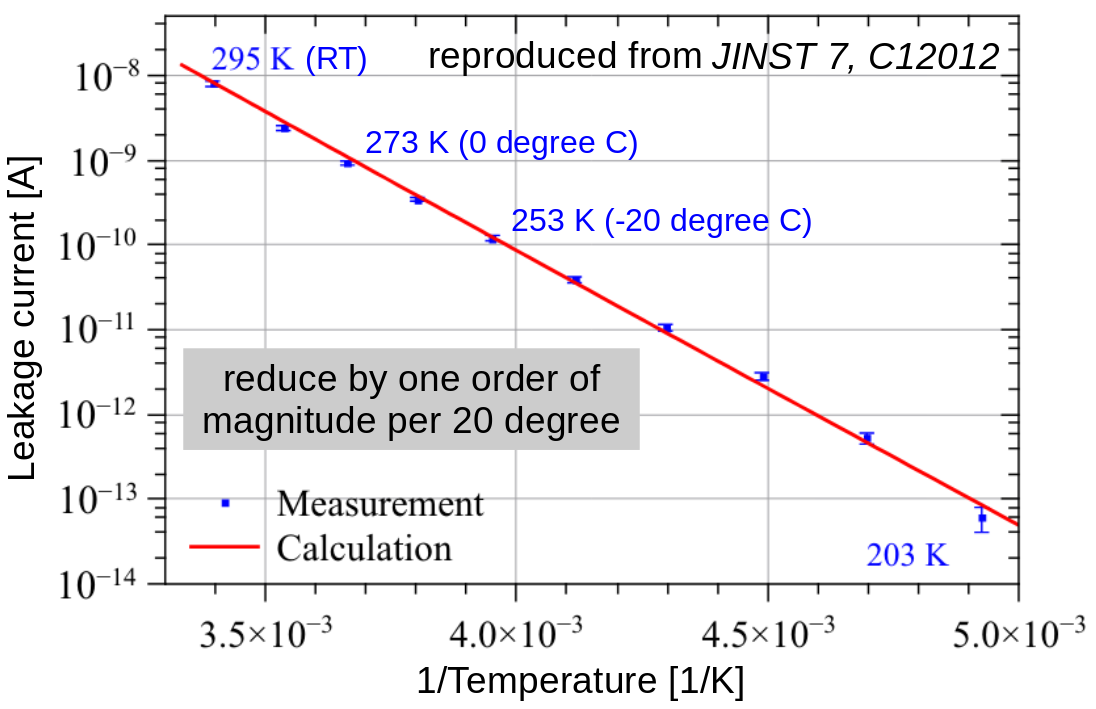}
(b)\includegraphics[height=50mm]{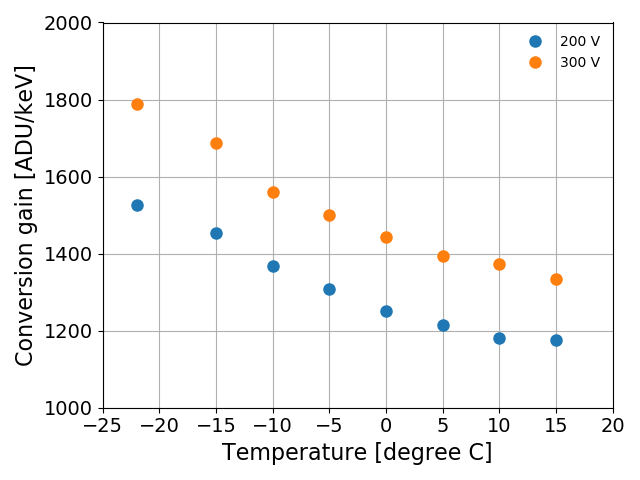}
\caption{Temperature dependence of (a) leakage current and (b) conversion gain. The leakage current is reproduced from ~\cite{Zhang2012}.}
\label{T_dependence}
\end{figure}

We have previously investigated the temperature dependence of the leakage current in the temperature range from 203~K to 295~K using a pad diode, as shown in figure~\ref{T_dependence}(a)~\cite{Zhang2012}. Lowering the operating temperature helps with the reduction of the leakage current: The leakage current can be reduced by one order of magnitude for a temperature difference of 20$^{\circ}$C, while the multiplication gain of the iLGAD sensor increases as well. Figure~\ref{T_dependence}(b) shows the measured conversion gain of the detector using 4.5~keV X-ray fluorescence as a function of the temperature between -22$^{\circ}$C and 15$^{\circ}$C at two different operating voltages (200 V and 300 V). The conversion gain in ADC unit per keV, which is proportional to the multiplication gain, increases by 29.2\% at 200~V and 33.3\% at 300~V for a temperature difference of 37$^{\circ}$C~\footnote{The multiplication gain of the iLGAD sensor at different temperatures can be determined through the ratio of conversion gains of the M\"onch detector using an iLGAD sensor and a conventional sensor.}.

\begin{figure}
\small
\centering
(a)\includegraphics[width=70mm]{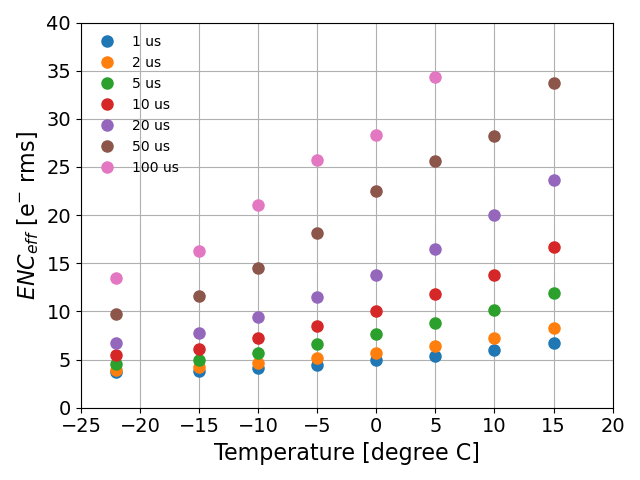}
(b)\includegraphics[width=70mm]{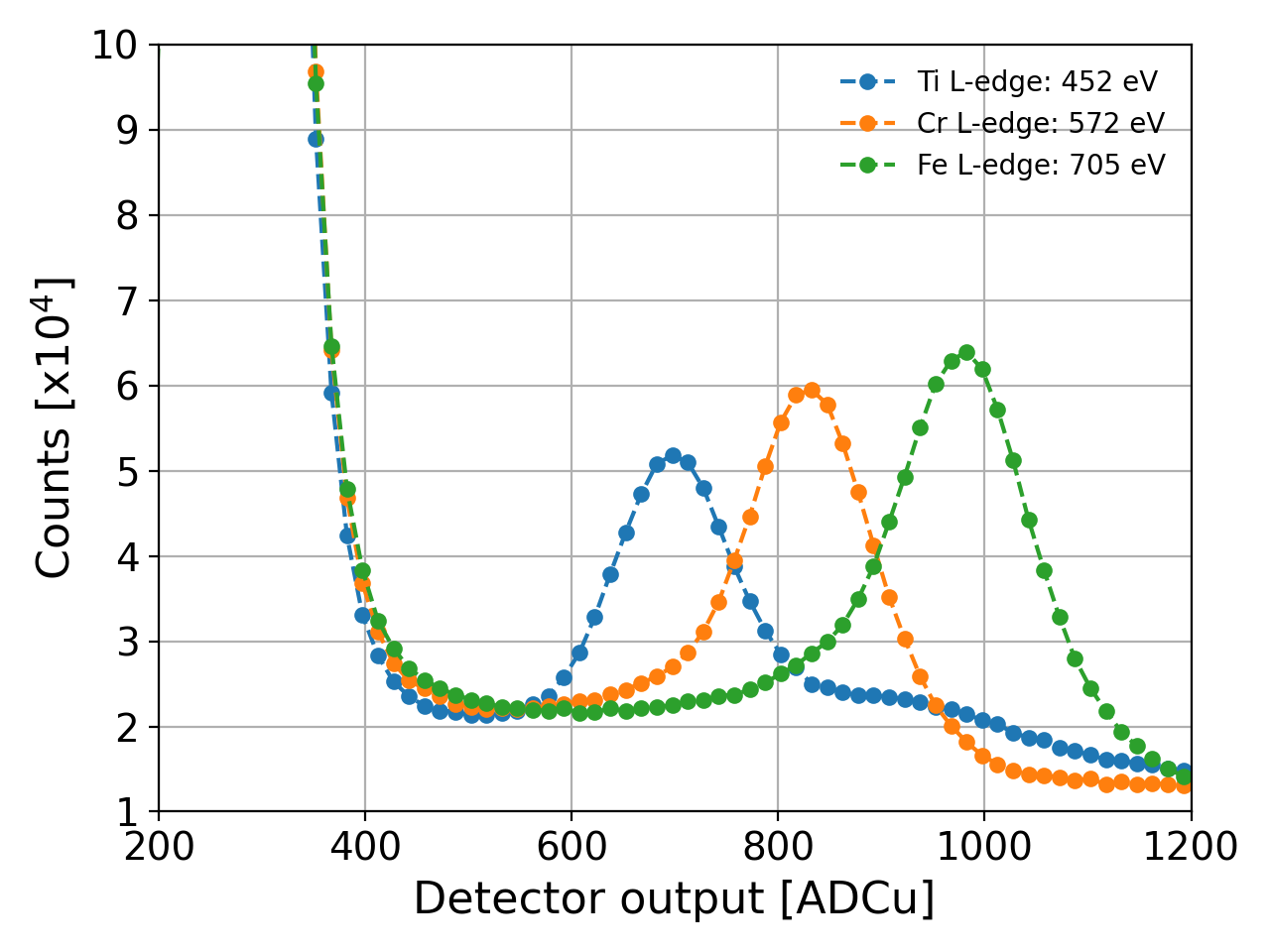}
\caption{(a) Effective Equivalent Noise Charge ($ENC_{eff}$) of the detector as a function of temperature for different exposure time. (b) Single photon detection for 452 eV, 572 eV and 705 eV X-ray fluorescence from the L-edges of Ti, Cr and Fe targets using an X-ray tube.}
\label{Spectra}
\end{figure}

%The effective Equivalent Noise Charge (ENC), $ENC_{eff}$, is defined as the $ENC$ ($rms$) divided by the multiplication gain $G$: $ENC_{eff} = ENC/G$.
The effective Equivalent Noise Charge (ENC), $ENC_{eff}$, is defined as:

\begin{equation}
\label{eq:noise}
  ENC_{eff} [e^{-}] = \frac{\sigma [ADU]}{gain [ADU/keV]} \cdot \frac{1000}{3.6 [eV]}
\end{equation}

\noindent with $\sigma$ the noise in ADC unit measured from a dark measurement, and $gain[ADU]$ the conversion gain. $ENC_{eff}$ has been measured as a function of temperature at 300~V and shown in figure~\ref{Spectra}(a) for different exposure time between  1~$\mu$s and 100~$\mu$s. An $ENC_{eff}$ of 3.53$\pm$0.18~e$^{-}$ $rms$ has been achieved at -22~$^{\circ}$C and 1~$\mu$s exposure time, when the shot noise due to the integrated leakage current is minimized. It corresponds to a SNR higher than 5 for soft X-rays above 63.5 eV (127.1 eV when clustering 2$\times$2 pixels). We demonstrate a reduction of the $ENC_{eff}$ by a factor of about 10 compared to the noise obtained with a conventional planar sensor with 1~$\mu$s exposure time at room temperature, which we can compare with the gain $\sim$ 11 for this sensor variation at this temperature.

In addition, measurements on the soft X-ray fluorescence from the L-edges of several metals have been performed at -2~$^{\circ}$C in vacuum. The iLGAD sensor was biased at 300~V and an exposure time of 10~$\mu$s was used in order to reduce the effective noise and at the same time to reach a good statistic. The spectra from the three fluorescence targets after clustering 2$\times$2 pixels in order to reduce the charge sharing effects between the 25~$\mu$m pitch pixels are shown in figure~\ref{Spectra}(b). The single photon peaks of 452~eV (in blue), 572~eV (in orange) and 705~eV (in green) can clearly be distinguished. This demonstrates single photon resolution at least down to 452~eV with a SNR of 23 (11.5 using 2$\times$2 pixel cluster). The lowest detectable photon energy under the same condition is expected to be $\leq$ 200~eV with a SNR greater than 5 using 2$\times$2 pixel clusters. However, this doesn't take into account the fact that at $\leq$ 200~eV most of the X-rays will be absorbed before entering the gain layer and the amplification will be lower. %Additional measurements at the SIM beamline of the Swiss Light Source using mono-energetic soft X-ray photons down to 200 eV will be done.

\begin{figure}
\small
\centering
(a)\includegraphics[width=70mm]{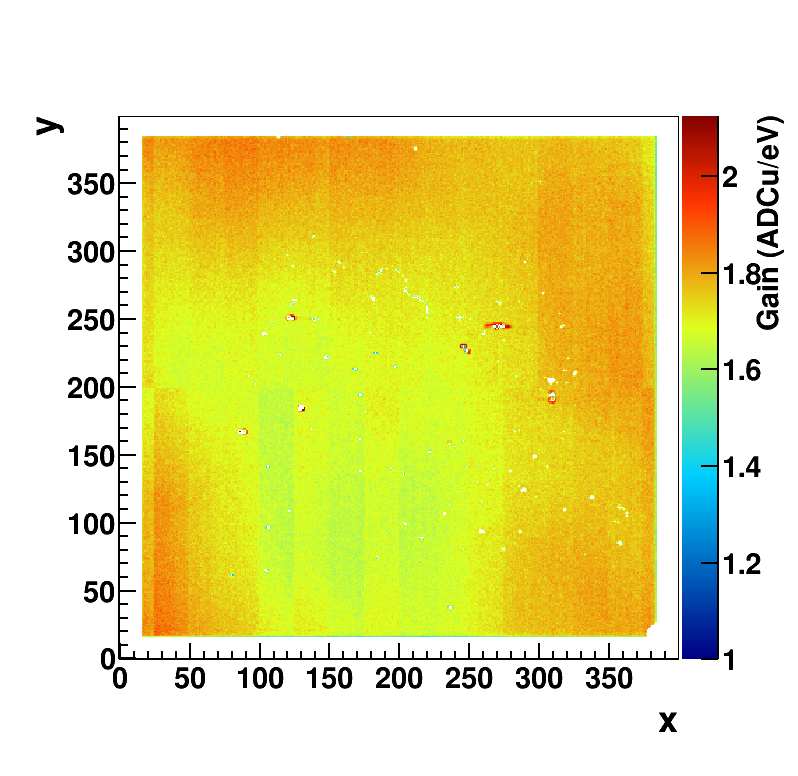}
(b)\includegraphics[width=70mm]{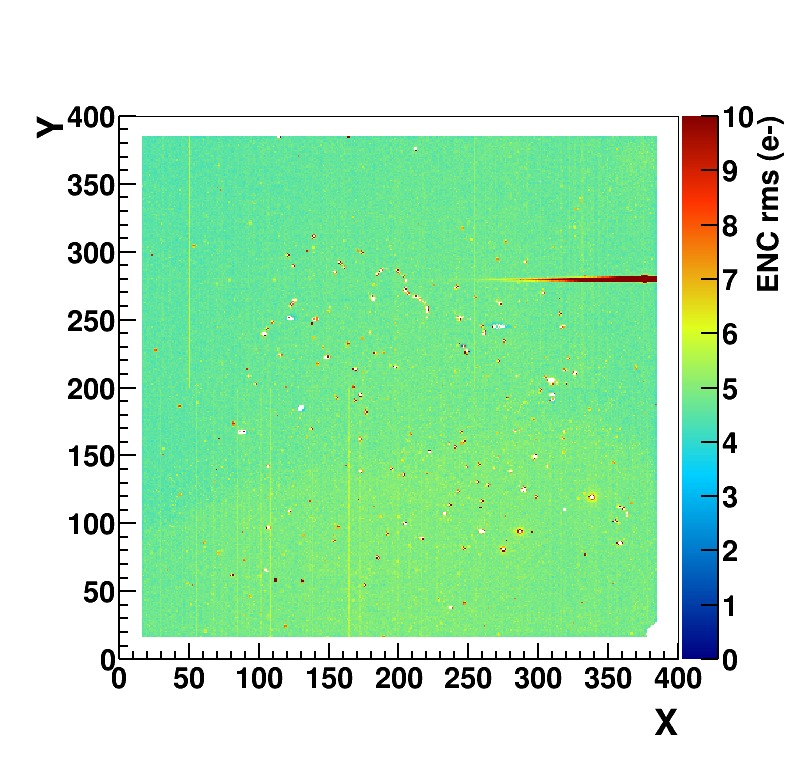}
\caption{The uniformity of the gain (a) and effective ENC over the entire sensor area. The first 9 pixels closest to the border are not connected to the readout ASIC and shown as blank in the image.}
\label{GainNoiseUniformity}
\end{figure}

Since the gain layer of the iLGAD sensor covers a larger detection area, the uniformity of the gain and noise is an important assessment. Figure~\ref{GainNoiseUniformity}(a) and (b) shows the gain and noise map for the investigated iLGAD sensor. The effective noise is distributed uniformly throughout the whole sensor; the gain map shows $\sim$3\% dispersion over the sensor area of 1$\times$1~cm$^{2}$ which is within an acceptable range. The gain dispersion is mainly caused by the non-uniformity of the multiplication gain, but part of it can also be attributed to the non-uniform temperature of the sensor and to the pixel-to-pixel variations. Since the gain correction has to be always performed for the charge-integrating detectors, the impact of the gain dispersion can be corrected.

\section{Summary and outlook}
\label{sec:summary}

The LGAD sensor with TEW are under development for soft X-ray detection using hybrid detectors. It involves the development of a TEW technology as well as the iLGAD technology optimized for soft X-rays. 

Simulations show that a TEW combining a shallow implant with a depth of a couple of hundred nm and a passivated surface is able to improve the QE up to 80\%. A batch of planar silicon sensor for the TEW optimization has been produced. Investigations for these sensors have been performed using either UV light of 405~nm or soft X-rays from 200~eV to 1250~eV and the experimental results of the QE measurements will be published in separate papers~\cite{Mar2022, Matteo2022}.

In addition, iLGAD sensors with the optimized TEW have been fabricated. First measurements were performed using the M\"onch readout ASIC. Single photon resolution at least down to 452~eV has been demonstrated and measurements on the lowest detectable photon energy have been planned using mono-energetic soft X-rays down to 200~eV at the SIM beamline of the SLS.

Due to the non-negligible leakage current of the iLGAD sensor due to its internal charge multiplication, these sensors require an operation with cooling in vacuum for soft X-rays and a short exposure time in order to reduce the effective noise and to maximize the SNR for single photon detection. These sensors already satisfy the requirements of FEL applications, where a short exposure time is required, while further optimization is necessary for synchrotrons where  a high duty cycle is required. We are investigating the performance of the iLGADs combined with single photon counting readout ASICs, which are less sensitive to the leakage current.

In the following development, the improvement of leakage current in iLGAD sensors will be emphasized. In addition, further optimization of the gain layer towards the surface will be explored for challenging the technological limit.

\acknowledgments

One of the authors, V. Hinger, has received funding from MSCA PSI-FELLOW-III-3i (EU grant agreement No. 884104).

%The authors would like to thank Thomas Huthwelker and Camelia Nicoleta Borca from the PHOENIX beamline at the SLS for their support on the investigations of the iLGAD sensors, Armin Kleibert and Joerg Raabe from the SIM beamline at the SLS for their help on the QE measurements with the planar silicon sensors with TEW.

%This is the most common positions for acknowledgments. A macro is available to maintain the same layout and spelling of the heading.

%\paragraph{Note added.} This is also a good position for notes added after the paper has been written.


\end{document}